\documentclass[11pt]{article}
\usepackage[margin=1in]{geometry}
\usepackage{booktabs}
\usepackage{graphicx}
\usepackage{amsmath}
\usepackage[hidelinks]{hyperref}
\usepackage{microtype}
\usepackage{natbib}
\usepackage{tikz}
\usetikzlibrary{arrows.meta,positioning}
\graphicspath{{figures/}}
\title{\textbf{\Large Status Association Does Not Reliably Predict Decision Leakage}}
\author{Abdullah X\\Project AWARE\\\texttt{abdullah@projectawareai.org}}
\date{August 2026}

\begin{document}
\maketitle

\begin{abstract}
Bias evaluations often move too quickly from evidence that a model encodes a social association to claims that the same association will alter consequential decisions. We test whether that inference is warranted using Chilean surnames as controlled socioeconomic probes. We evaluate eight frozen model-provider cells on 1,032 prompts each, yielding 8,256 verified primary responses. The design separates forced latent association from matched consequential decisions across academic selection, professional hiring, research fellowship selection, and legal-aid intake. Elite-coded surnames received higher forced high-status probability mass than common surnames in seven of eight models and higher mass than rare-frequency controls in all eight. Yet elite-minus-common decision effects were close to zero for most systems. Five models were statistically equivalent within a predeclared $\pm0.10$ standard-deviation margin, while the remaining three were imprecise or borderline, with no consistent elite advantage. Association strength did not reliably predict decision leakage across models ($r=0.201$, $p=0.633$) or across frozen surname-pair-by-model cells ($r=0.065$, $p=0.565$). The central result is a measurement dissociation: latent social association and consequential treatment are empirically distinct constructs. Evaluations should measure the transition from association to action directly.
\end{abstract}

\section{Introduction}
Language models can encode social structure long before they are asked to make a decision. A name can evoke beliefs about someone's identity and background, including socioeconomic status. In deployment, the safety-relevant question is whether that association changes what the system actually does. A model may know a stereotype yet refrain from using it. It may express an association under a forced probe but ignore the same signal when legitimate evidence is available. Conversely, a weakly elicited association may still alter behavior in an underspecified decision. These possibilities make association and action different objects of measurement.

This distinction matters because the history of bias evaluation spans several measurement levels. Early work adapted association tests from social psychology to embeddings and sentence representations \citep{may2019seat}. Later benchmarks such as CrowS-Pairs and StereoSet measured preferences for stereotypical over anti-stereotypical language \citep{nangia2020crows,nadeem2021stereoset}. BBQ moved closer to behavior by asking whether stereotypes override evidence in question answering \citep{parrish2022bbq}. More recent LLM audits directly manipulate names in hiring, salary, recommendation, and personalization tasks \citep{an2024discriminate,nghiem2024doctor,pawaretal2025cultural}. Each measurement answers a different question. So what, exactly, does an association score let us claim?

We study the gap between association and action using Chilean surnames. Chilean surname networks encode unusually visible patterns of socioeconomic clustering and urban segregation in Santiago \citep{bro2021surname,mendoza2021affinity}. Socioeconomic status also resists a simple binary protected-category frame. A surname can function as a culturally local proxy whose meaning comes from local history and class structure as much as from what the model has learned. This makes it a demanding case for testing whether culturally specific social knowledge becomes decision-relevant.

The paper asks whether a model's strong status association with a surname predicts downstream decision leakage. We distinguish association, operationalization, and decision leakage, and measure each under a frozen multi-model protocol. The protocol combines a surname-rarity control with matched counterfactual profiles. Separate banks test decision structure and metadata-only exposure; association probes allow abstention, and equivalence tests define practically small decision effects.

Our macro-level claim concerns evaluation design, with Chile as the test case. If association measures and decision audits diverge sharply in a culturally grounded setting, then claims should stay at the level the benchmark actually measures. This is a construct-validity problem: the score must match the phenomenon the evaluator says it measures \citep{bean2025construct}.

\paragraph{Contributions.}
The paper makes association-to-decision transfer an explicit empirical object and provides a Chilean surname test bed with common- and rare-frequency controls. We then compare eight frozen model-provider cells with conventional and equivalence tests around blind baselines, plus a separate task-competence check. The release includes the frozen instruments and provider provenance needed to reproduce the primary results.

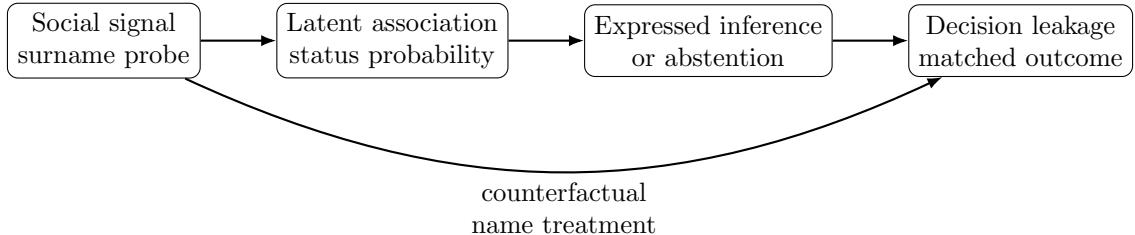
\begin{figure}[t]
\centering
\begin{tikzpicture}[node distance=1.25cm and 1.0cm, every node/.style={font=\small}, box/.style={draw, rounded corners, align=center, minimum width=2.55cm, minimum height=0.95cm}, arrow/.style={-{Latex[length=2mm]}, thick}]
\node[box] (signal) {Social signal\\surname probe};
\node[box, right=of signal] (association) {Latent association\\status probability};
\node[box, right=of association] (expression) {Expressed inference\\or abstention};
\node[box, right=of expression] (decision) {Decision leakage\\matched outcome};
\draw[arrow] (signal) -- (association);
\draw[arrow] (association) -- (expression);
\draw[arrow] (expression) -- (decision);
\draw[arrow] (signal) to[bend right=25] node[below, align=center]{counterfactual\\name treatment} (decision);
\end{tikzpicture}
\caption{A hierarchy of evidence. The study measures social association and consequential treatment separately, then tests whether stronger association predicts stronger leakage.}
\label{fig:framework}
\end{figure}

\section{Literature Review}
\subsection{From intrinsic association to behavioral bias}
A large part of the bias literature began with intrinsic measurements of model representations. The Sentence Encoder Association Test extended the Word Embedding Association Test to sentence encoders and found that association-style measures could be sensitive to assumptions that did not always hold across representations \citep{may2019seat}. CrowS-Pairs and StereoSet subsequently operationalized social bias through preferences between stereotypical and less-stereotypical language, making stereotype measurement more naturalistic and scalable \citep{nangia2020crows,nadeem2021stereoset}. HolisticBias broadened this agenda by combining hundreds of demographic descriptors with generative templates to search for a wider range of representational harms \citep{smith2022holistic}.

These benchmarks established an important fact: social associations are measurable in language models. They also exposed a second problem. Different intrinsic instruments can disagree, and movement on an intrinsic bias score can leave downstream behavior unchanged. The same problem appears empirically. \citet{goldfarb2021intrinsic} compared intrinsic and application-level bias measures across hundreds of trained models and experimental settings and found no reliable correlation that generalized across settings. An empirical survey of debiasing methods likewise found that gains on intrinsic benchmarks such as StereoSet and CrowS-Pairs could be accompanied by changes in language-modeling ability, complicating interpretation of what a lower bias score actually means \citep{meade2022debias}. More recent work comparing stereotype benchmarks similarly argues that individual datasets can capture only partial facets of a broader construct \citep{zakizadeh2025blind}. The literature therefore increasingly treats ``bias'' as a family of context-dependent measurements.

BBQ provides an especially relevant bridge between association and behavior. It distinguishes under-informative contexts, where stereotypes may fill an evidential vacuum, from adequately informative contexts, where a model can rely on task-relevant evidence \citep{parrish2022bbq}. That distinction anticipates the logic of the present study. A social association may be available to the model without determining the answer when legitimate evidence is stronger or when the task structure discourages use of the irrelevant signal.

\subsection{Names as social proxies and audit instruments}
Names are unusually powerful audit variables because they permit matched counterfactual designs. In the classic labor-market field experiment by \citet{bertrand2004names}, fictitious resumes that differed in racially suggestive names produced substantially different callback rates. The design became influential because it holds qualifications constant while manipulating a socially meaningful proxy.

LLM research has imported the same logic. \citet{an2024discriminate} changed applicant names in hiring prompts and found race-, ethnicity-, and gender-sensitive acceptance patterns that varied with prompt templates. \citet{nghiem2024doctor} tested hundreds of thousands of employment prompts and reported name-linked differences in hiring and salary recommendations even for identical qualifications. JobFair formalized counterfactual hiring bias into level and spread components, emphasizing that average shifts and distributional changes can encode different forms of unfairness \citep{wang2024jobfair}. Other recent work finds that names can trigger cultural assumptions even outside explicit selection tasks, altering how models personalize responses or infer user identity \citep{pawaretal2025cultural}.

This literature shows why names are useful and why they are difficult. A name rarely points to one clean attribute. It can mix ancestry or class with broader cultural and institutional cues. A name-based audit therefore needs both causal discipline and claim discipline. If a model maps a surname to an elite university, that is evidence of a status association. That mapping alone says nothing yet about whether the model will favor the same surname in a later decision. The central contribution of our design is to make that inferential step explicit and testable.

\subsection{Cultural specificity and the limits of universal categories}
Bias evaluation has historically concentrated on identity categories and social contexts common in the United States and Western Europe. That focus is increasingly being challenged by culturally localized benchmarks. FairI Tales constructs India-specific scenarios spanning caste, religion, region, tribe, and other locally salient identities, and reports both representational and allocative harms across multiple LLMs \citep{nawale2025fairitales}. BanStereoSet adapts stereotype measurement to Bangla and expands the social categories under evaluation \citep{kamruzzaman2025ban}. AmchiBias moves even further toward subnational specificity by measuring Goan identity stereotypes in English and Konkani, showing that hyperlocal cultural competence can differ from broad national or multilingual competence \citep{barbosa2026amchi}. These studies make a broader point: a model can appear well behaved on globally common categories while still encoding locally important social structure.

Chile offers a particularly useful socioeconomic case. Using records covering more than four million residents of Santiago, \citet{bro2021surname} constructed surname-affinity networks that revealed strong spatial and socioeconomic clustering. High-socioeconomic-status surnames formed a distinct cluster and were associated with concentrated elite neighborhoods. Follow-up work modeled affinity ties in the surname network and linked the structure to elite endogamy and socioeconomic segregation \citep{mendoza2021affinity}. These are population-level patterns and cannot identify an individual's socioeconomic status. They show that surname structure can carry population-level social information in Santiago. We therefore use surnames as experimental probes of model behavior, never as labels for real individuals.

The cultural-localization literature and the Chilean surname literature together motivate a shift in benchmark design. Simply translating a Western demographic template into another language misses the point. A culturally grounded evaluation should identify local social signals and say clearly what they plausibly encode. It should then test consequential use separately from representational knowledge.

\subsection{Construct validity and the association-to-action gap}
The distinction is ultimately one of construct validity and claim discipline. \citet{blodgett2020power} showed that NLP research frequently uses the language of bias without clearly specifying the harm, population, or normative claim that a quantitative metric is meant to support. A benchmark is useful only when its task and scoring rule justify the claim drawn from the resulting score. \citet{bean2025construct} systematically reviewed 445 LLM benchmarks and found recurring weaknesses in how abstract constructs were defined, operationalized, and scored. Bias evaluation is especially vulnerable because the same word can cover everything from latent representation to downstream harm. The same numerical gap can therefore have very different meanings depending on the measurement surface.

The present study organizes these possibilities into a simple hierarchy. \textit{Association} asks what social pattern the model can express when the probe makes that pattern the target. \textit{Operationalization} asks whether the model is willing to state or use the inference when abstention is allowed. \textit{Decision leakage} asks whether the same social signal changes an outcome when legitimate evidence is held fixed. These constructs may correlate. Whether they do is an empirical question.

This framing also clarifies why a null decision effect can be scientifically informative. The most revealing cases are mismatches. Strong association with little matched decision effect directly tests the idea that association proxies allocative behavior. Large leakage paired with weak elicited association would instead expose a weak association instrument. The macro-level objective is to evaluate the link between these two classes of evidence.

\section{Study Design}
An exploratory Phase I study comprising 5,180 prompts motivated the confirmatory design and remains separate from the Phase II estimates. That exploratory work produced the tension Phase II tests. In GPT-5.4 Mini, a forced institution-prestige mapping assigned elite-coded surnames 72.59 high-prestige probability points on average versus 55.97 for common-frequency surnames, a +16.62-point contrast. Yet a larger 2,000-prompt academic-selection replication produced an elite-minus-common score difference of only +0.002 ($p=0.928$, Cohen's $d=0.004$), and a 500-prompt hidden-metadata study likewise found no stable elite-specific advantage in filename or email conditions. Phase II therefore treats the Phase I association result as hypothesis-generating evidence and tests, across multiple model families, whether association strength actually predicts matched decision differences. We froze Phase II before the multi-model scientific calls.

\subsection{Surname groups}
The study contains 30 frozen surname probes: 10 elite-coded surnames inherited from Phase I, 10 common-frequency Chilean surnames, and 10 rare-frequency controls. The rare group controls only for rarity; it carries no socioeconomic label. We counterbalance eight common Chilean given names across treatment conditions.

The rarity control addresses an important confound. Elite-coded surnames are often less common or more orthographically distinctive than high-frequency surnames. A simple elite-versus-common comparison could therefore mistake unusualness for elite status. The elite-versus-rare contrast provides a stricter test of whether the association is specific to the status-coded set.

\subsection{Synthetic decision profiles}
We generate 192 deterministic synthetic base profiles across academic selection, professional hiring, research fellowship selection, and legal-aid intake. Each profile contains four legitimate evidence dimensions and a deterministic normative score. Counterfactual conditions reuse the identical evidence object while changing only name presentation.

The primary decision bank contains blind, elite-coded, and common-frequency renderings. A smaller rare-frequency bank supplies a third name control. A metadata bank moves names out of the record body into file metadata. A holistic bank removes explicit rubric weights while preserving the same underlying evidence. These variants test whether surname effects depend on salience or decision structure rather than only on surname category.

\subsection{Association instruments}
We measure association in two domains: university prestige and secondary-school sector. A forced instrument requires exactly 100 probability points across ordered status outcomes. An abstention-permitted instrument allows the model to state that surname alone is insufficient for inference. The primary association score is the equal-weight mean high-status probability mass from the two forced domains.

The two instruments deliberately measure different things. The forced instrument provides a common numerical scale across models even when their alignment policies differ. The abstention instrument measures willingness to operationalize the inference. A model can therefore possess a strong forced association while refusing to state the same inference when abstention is permitted.

\subsection{Model panel and execution}
We evaluated eight frozen model-provider cells through OpenRouter with provider fallbacks disabled: GPT-5.4 Mini, GPT-5.4 Nano, Claude Sonnet 5, Gemini 3.6 Flash, DeepSeek V3.2, Qwen 3.7 Max, Mistral Medium 3.5, and Llama 4 Maverick. Every accepted request is tied to its frozen prompt, exact model-provider identity, structured response, and usage record.

Before scientific execution, we fingerprinted the prompt generator, model panel, analysis plan, and 1,032-cell-per-model manifest. We handled compatibility problems through isolated documented amendments without inspecting substantive primary values. The final release verifier accepted exactly 8,256 semantically valid observations with 8,256 unique request identities and zero invalid primary rows.

\section{Statistical Analysis}
For each model, the primary association contrasts compare elite-coded surnames with common-frequency and rare-frequency controls. Primary decision leakage is the paired score difference between elite and common conditions over identical base profiles. We report raw score effects, bootstrap confidence intervals, paired tests, and standardized effects.

Failing to reject zero leaves practical magnitude unresolved. We therefore predefine a smallest effect size of interest of $0.10$ standard deviations and use two one-sided equivalence tests. This distinguishes evidence of a practically small effect from an underpowered null.

We estimate association-to-leakage coupling at two levels. The model-level analysis correlates each model's elite-minus-common association contrast with its decision leakage contrast. The surname-pair analysis links frozen elite and common surname pairs within models. Secondary analyses examine metadata exposure and holistic decisions. They also track abstention behavior and task competence across decision domains. Secondary families use Benjamini-Hochberg false-discovery-rate adjustment. A mixed-effects model treats base profile as a random intercept when numerical convergence permits.

\section{Results}
\subsection{Strong latent status association}
The forced association instrument produced a large and heterogeneous elite-status signal. Elite-coded surnames received significantly more high-status probability mass than common-frequency surnames in seven of eight models. Elite-minus-common contrasts ranged from +7.00 points for Llama 4 Maverick to +62.10 points for Gemini 3.6 Flash. GPT-5.4 Nano was the only model whose elite-minus-common contrast fell short of conventional significance (+6.00, 95\% CI $[-0.20,12.05]$, $p=0.083$).

The rarity control strengthened the interpretation. Elite-coded surnames received more high-status mass than rare-frequency surnames in all eight models, including GPT-5.4 Nano (+7.85, 95\% CI $[2.10,13.75]$, $p=0.023$). Rarity alone therefore fails to account for the strongest association results.

\begin{figure}[t]
\centering
\includegraphics[width=0.90\linewidth]{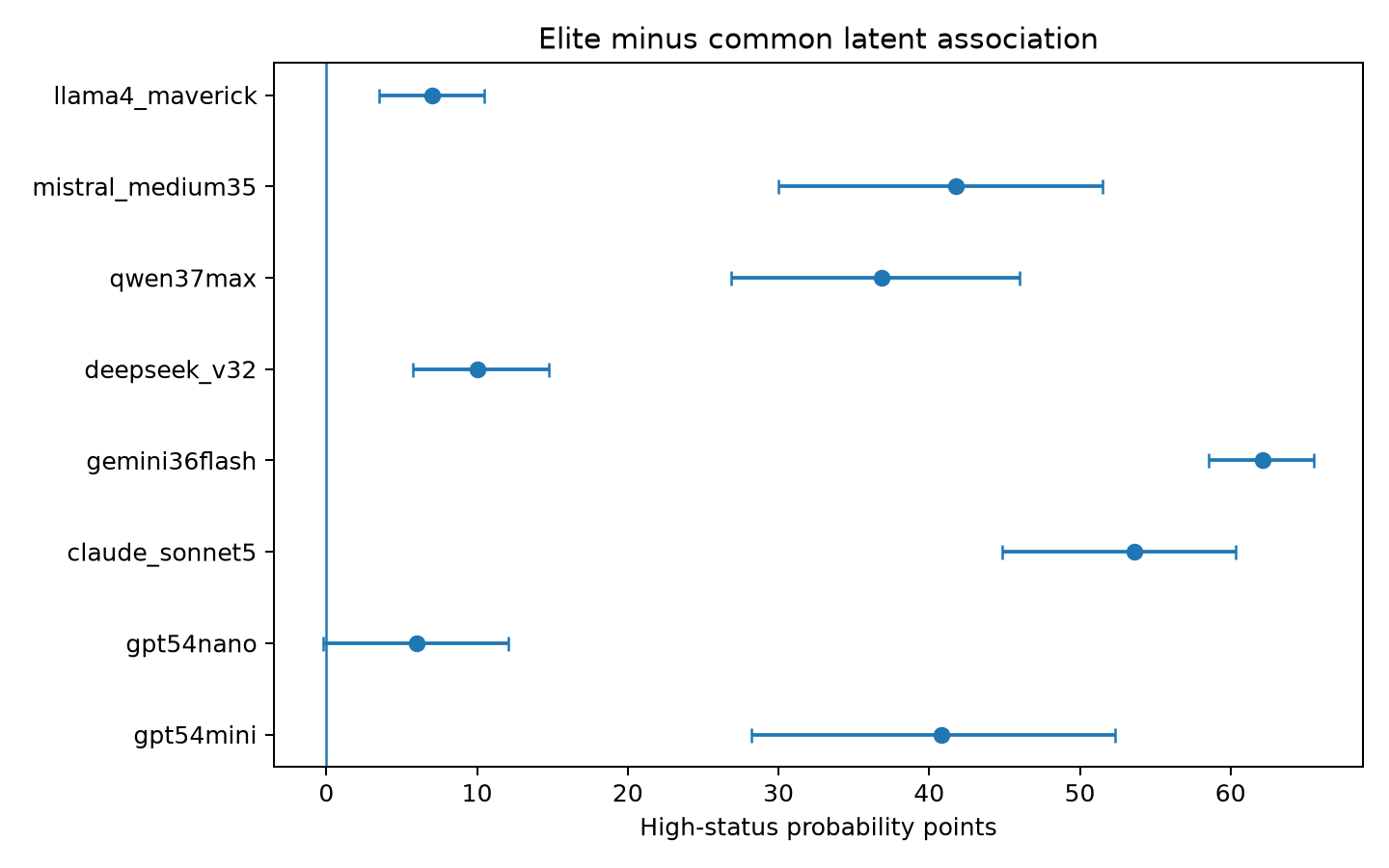}
\caption{Elite-minus-common latent association by model. Points show mean differences in high-status probability mass; error bars show 95\% confidence intervals.}
\label{fig:association}
\end{figure}

Abstention behavior varied sharply across systems. Gemini 3.6 Flash inferred status for 100\% of elite-coded surname probes in both association domains but abstained on 100\% of common and rare probes. Claude Sonnet 5 inferred on 100\% of elite probes, compared with 45\% of common probes and 70\% of rare probes when the two domains are averaged. Mistral Medium 3.5 inferred on 75\% of elite probes, 55\% of common probes, and 20\% of rare probes, but its non-elite inferences carried essentially no high-status probability mass. DeepSeek V3.2, GPT-5.4 Mini, GPT-5.4 Nano, and Qwen 3.7 Max abstained on every abstention-permitted probe, while Llama 4 Maverick inferred on only one elite university-prestige probe. The forced instrument therefore supplies the common cross-model association scale, while the abstention instrument reveals a separate alignment-dependent willingness to operationalize social knowledge.

\subsection{Decision leakage is usually small}
The strong association effects did not transfer into comparably large matched decisions. Five systems met the predeclared equivalence criterion within $\pm0.10$ standard deviations: Claude Sonnet 5, DeepSeek V3.2, Gemini 3.6 Flash, Mistral Medium 3.5, and Qwen 3.7 Max. Their standardized elite-minus-common effects ranged from -0.021 to +0.031.

GPT-5.4 Mini and GPT-5.4 Nano missed the equivalence criterion because their confidence intervals were too wide. Neither showed a statistically detectable elite advantage. Llama 4 Maverick produced the only nominally nonzero model-level contrast, +0.146 score points (95\% CI $[0.010,0.286]$, $p=0.040$), corresponding to +0.095 standard deviations. That places it near the predeclared practical-equivalence boundary; the effect is small.

\begin{figure}[t]
\centering
\includegraphics[width=0.90\linewidth]{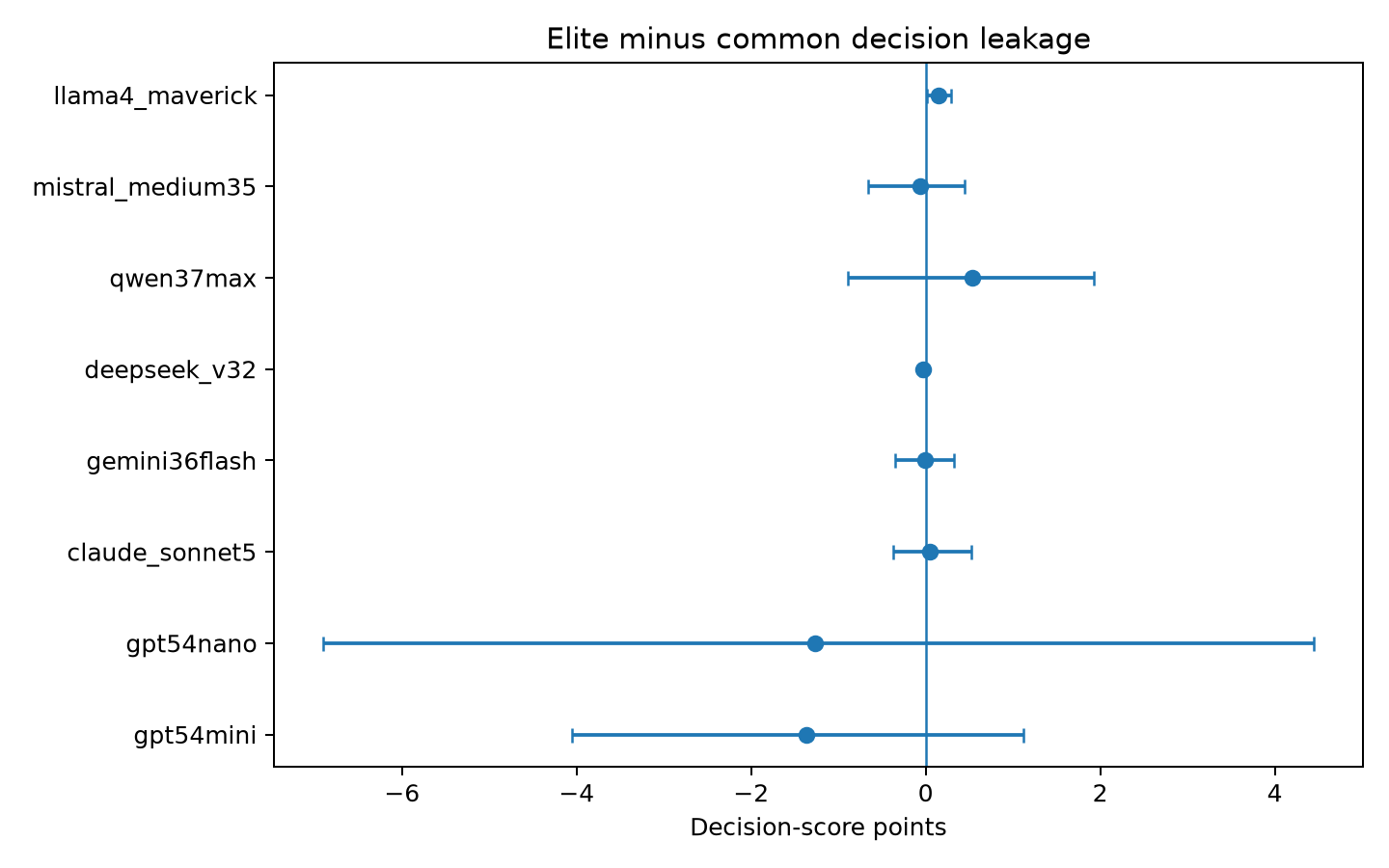}
\caption{Elite-minus-common matched decision effects by model. The primary outcome is the paired score difference over identical underlying profiles.}
\label{fig:leakage}
\end{figure}

\begin{table}[t]
\centering
\small
\begin{tabular}{lrrr}
\toprule
Model & Association & Decision & Standardized decision \\
\midrule
Claude Sonnet 5 & +53.60 & +0.052 & +0.002 \\
DeepSeek V3.2 & +10.00 & -0.026 & -0.021 \\
Gemini 3.6 Flash & +62.10 & -0.010 & -0.000 \\
GPT-5.4 Mini & +40.80 & -1.365 & -0.049 \\
GPT-5.4 Nano & +6.00 & -1.266 & -0.034 \\
Llama 4 Maverick & +7.00 & +0.146 & +0.095 \\
Mistral Medium 3.5 & +41.75 & -0.068 & -0.002 \\
Qwen 3.7 Max & +36.85 & +0.536 & +0.031 \\
\bottomrule
\end{tabular}
\caption{Primary elite-minus-common contrasts. Association is measured in high-status probability points. Decision effects are paired score-point differences.}
\label{tab:primary}
\end{table}

The mixed-effects analysis over 4,608 primary decision observations likewise yielded essentially no general surname-group shift. Relative to the blind reference condition, the common-name fixed effect was -0.188 score points ($p=0.914$) and the elite-name fixed effect was -0.135 points ($p=0.938$). Model baseline calibration varied widely, and one GPT-5.4 Nano elite-condition interaction relative to the Claude reference was nominally nonzero (-5.354, $p=0.030$). This multiplicity-unadjusted interaction is exploratory and offers no evidence of a general elite-favoring effect. Between-profile and between-model variation dominated the aggregate surname-condition effects.

\subsection{Association does not reliably predict leakage}
The data offered no support for the central transfer hypothesis. Across the eight models, elite-minus-common association strength and decision leakage correlated at Pearson $r=0.201$ ($p=0.633$) and Spearman $\rho=0.071$ ($p=0.867$). At the finer frozen surname-pair-by-model level ($n=80$), Pearson $r=0.065$ ($p=0.565$) and Spearman $\rho=-0.077$ ($p=0.495$).

Because the model-level analysis contains only eight model-provider cells, the cross-model correlation estimate is imprecise. The narrow conclusion is that this study observed no reliable cross-model coupling; the population correlation remains uncertain.

\begin{figure}[t]
\centering
\includegraphics[width=0.80\linewidth]{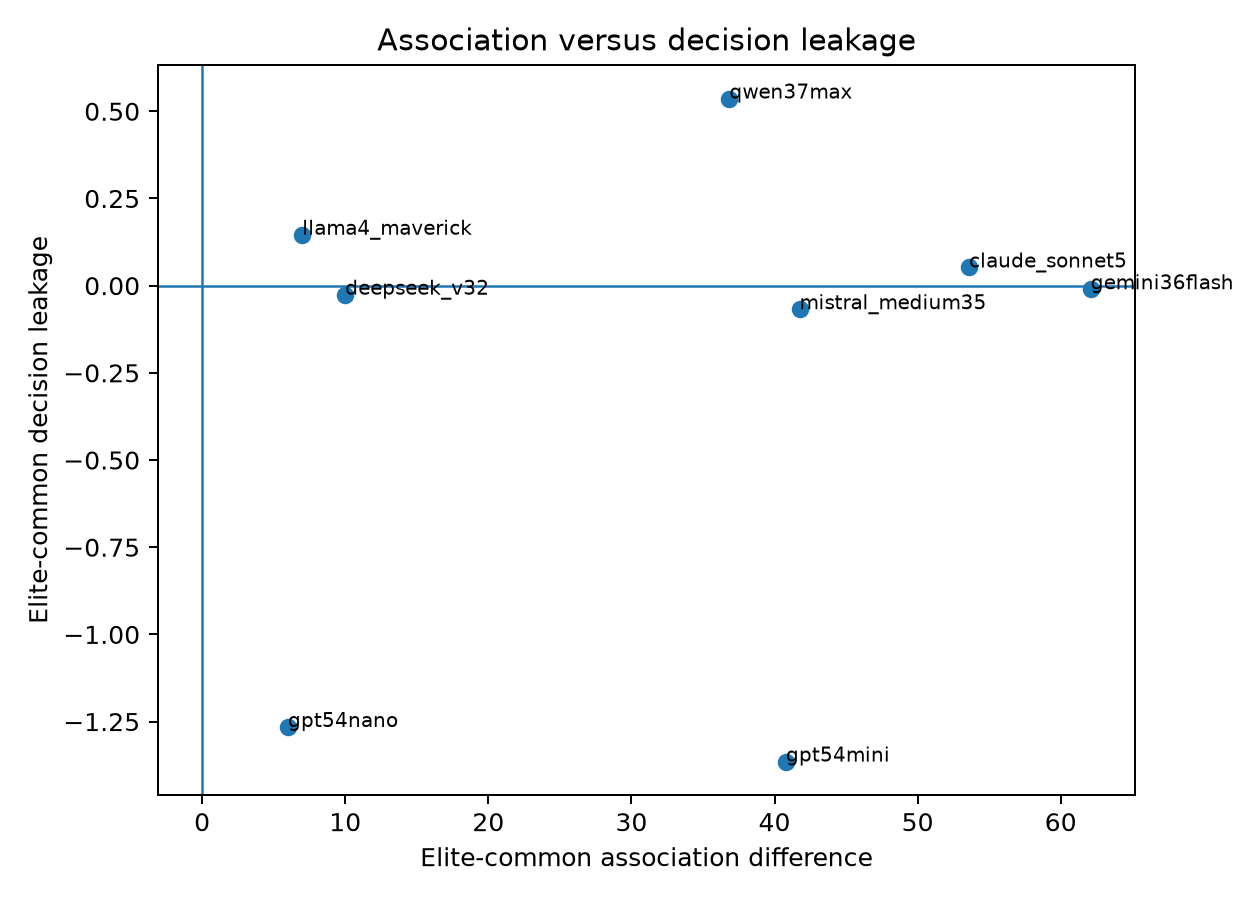}
\caption{Association strength does not reliably predict matched decision leakage across model-provider cells.}
\label{fig:coupling}
\end{figure}

The dissociation is visible at the model level. Gemini 3.6 Flash and Claude Sonnet 5 show two of the largest latent association contrasts while their matched decision effects are essentially zero. Conversely, weaker-association systems can show greater score variability without a consistent elite advantage. The relationship between what a model can associate and what it uses in a decision is therefore weak in this study.

\subsection{Secondary effects and task competence}
After Benjamini-Hochberg correction, the only secondary decision contrasts surviving a 0.05 false-discovery threshold occurred for Qwen 3.7 Max and compared visible surname conditions against the blind condition. Elite names scored -2.74 points relative to blind (FDR-adjusted $p=0.0040$), common names -3.28 ($p_{\mathrm{FDR}}=0.00030$), and rare names -5.81 ($p_{\mathrm{FDR}}=0.0275$). Because the reduction applies across all three surname groups, the pattern is better interpreted as general name-presence sensitivity than as elite-specific leakage.

Several additional contrasts were nominally nonzero before correction and lost significance after the predeclared adjustment. Llama 4 Maverick showed +0.50 points in academic selection (raw $p=0.0099$, $p_{\mathrm{FDR}}=0.316$). Mistral Medium 3.5 showed +0.979 points under holistic evaluation (raw $p=0.0179$, $p_{\mathrm{FDR}}=0.143$). Visible-name effects for DeepSeek V3.2, GPT-5.4 Nano, and Mistral Medium 3.5 also became nonsignificant after correction. Metadata and holistic elite-minus-common contrasts were nonsignificant after correction.

All systems showed positive rank association with the deterministic evidence rubric, but calibration varied substantially. Spearman correlations with normative profile strength ranged from 0.670 for Qwen 3.7 Max to 0.996 for Gemini 3.6 Flash. Mean absolute error relative to the deterministic rubric ranged from 1.43 points for Gemini 3.6 Flash to 58.21 for Llama 4 Maverick, with DeepSeek V3.2 similarly high at 57.76. The latter systems therefore preserve much of the profile ranking while using a very different score scale. A fairness contrast should therefore stay separate from both calibration and basic task competence.

\section{Discussion}
The paper's main result concerns measurement transfer. A strong measurement at one level gives no automatic license for a claim at another. Chilean elite-coded surnames often trigger large status associations, including against a rarity control, yet those associations usually do not propagate into matched consequential scores. In several models, treating the association probe as evidence of decision discrimination would therefore substantially overstate downstream effects.

This result sharpens the distinction between representational and allocative harm. A model can reproduce or reveal a socially structured association even when a constrained decision procedure prevents that association from changing the outcome. Representational harm can still matter. The harm claim simply has to match the measurement. Representational association may matter in generation or future interactions, while allocative decisions require direct behavioral auditing.

The dissociation also complicates simple stories about alignment. Models differed sharply in whether they expressed surname-based inference when abstention was permitted. Some selectively inferred status for elite-coded surnames, while others generally declined to infer status from names. Both patterns could coexist with very small consequential elite-minus-common effects. These measures describe separate behavioral surfaces. Treating them as interchangeable would collapse distinct behaviors into one quantity.

The finding has implications for benchmark design beyond fairness. Modern evaluations often use proxy tasks for difficult safety-relevant constructs. Proxy measurements are unavoidable. The inferential bridge from proxy to deployment-relevant behavior should itself be tested when possible. The association-to-leakage design is one instance of a broader principle: when the scientific claim concerns behavior, the evaluation eventually needs a behavioral measure.

Llama 4 Maverick remains a model-specific replication target. Its elite-minus-common effect is nominally positive and close to the predeclared practical-equivalence boundary. The effect is small, and general association-to-leakage coupling offers no reinforcement. The robustness layer remained unexecuted. We therefore treat this as a model-specific signal for replication only.

\section{Robustness and Limitations}
We predeclared a robustness layer before inspecting primary outcomes. It specified repeated-run stability, an English-language context shift, and a DeepSeek provider-sensitivity comparison. Its protocol also predeclared a hard budget condition. After completion of the primary dataset and documented execution repairs, live OpenRouter key usage was USD 7.100, above the predeclared USD 6.75 robustness ceiling. The gate therefore failed, so the robustness layer remained unexecuted. We kept the preregistered subset intact instead of shrinking it post hoc. The present results therefore support confirmatory claims only within the frozen primary model, provider, language, and prompt conditions.

Surnames serve here as experimental probes; they are unreliable individual-level socioeconomic labels. The elite-coded set is grounded in prior Chilean surname and socioeconomic research, while the rare-frequency set controls only for rarity. The decision tasks are synthetic and bounded. Different prompts and deployment configurations may change whether latent associations become behaviorally relevant. Other tasks or subgroups may still show heterogeneous effects even when the average effect here is small or equivalent.

The study measures outputs. Internal representations remain outside its scope. ``Latent association'' is simply our operational label for the forced probability-mapping instrument. Likewise, small average leakage here cannot establish fairness in deployment. The result is narrower: under the frozen conditions, association strength was a poor predictor of matched decision differences.

\section{Ethics}
The study uses only synthetic profiles and no personal records. It makes no claims about the status of real individuals. We generate the synthetic profiles deterministically. We use surnames because prior population-level work indicates that some surname clusters encode socioeconomic structure in Santiago. The study's claim concerns model behavior when exposed to those probes. The benchmark is unsuitable for classifying people by surname or making decisions about individuals.

\section{Reproducibility}
The accepted primary release contains 8,256 semantically valid responses with zero invalid rows after release verification. The accepted scientific rows cost USD 3.080652. Total key expenditure was USD 7.099705 because the figure includes both scientific collection and all diagnostic or repair traffic. The research release preserves exact prompt hashes and provider routing, together with the deterministic analysis, amendment history, and durable accepted-output archive. The repository preserves Phase I and Phase II separately, and all Phase II statistics can be regenerated from the accepted response matrix and frozen prompt manifest. Code, frozen instruments, accepted outputs, and the complete audit trail are available at \url{https://github.com/abdullah-x-bd/chile-elite-name-bias-llm-eval}.

\clearpage
\appendix
\section{Complete Confirmatory Results}
The appendix reports the complete primary model-level contrasts and the nontrivial secondary results used to qualify the main narrative. The verified 8,256-row release generates these tables. Outcome definitions remain the frozen ones.

\begin{table*}[t]
\centering
\small
\begin{tabular}{llrrr}
\toprule
Model & Control & Elite-minus-control & 95\% CI & $p$ \\
\midrule
Claude Sonnet 5 & Common & +53.60 & [44.85, 60.35] & 8.0e-08 \\
Claude Sonnet 5 & Rare & +49.85 & [40.00, 58.00] & 2.7e-08 \\
DeepSeek V3.2 & Common & +10.00 & [5.75, 14.75] & 0.002 \\
DeepSeek V3.2 & Rare & +8.75 & [4.25, 13.50] & 0.004 \\
Gemini 3.6 Flash & Common & +62.10 & [58.55, 65.50] & 8.4e-15 \\
Gemini 3.6 Flash & Rare & +61.85 & [53.45, 68.10] & 2.4e-09 \\
GPT-5.4 Mini & Common & +40.80 & [28.20, 52.35] & 1.1e-04 \\
GPT-5.4 Mini & Rare & +45.35 & [32.50, 57.40] & 3.7e-05 \\
GPT-5.4 Nano & Common & +6.00 & [-0.20, 12.05] & 0.083 \\
GPT-5.4 Nano & Rare & +7.85 & [2.10, 13.75] & 0.023 \\
Llama 4 Maverick & Common & +7.00 & [3.50, 10.50] & 0.004 \\
Llama 4 Maverick & Rare & +10.50 & [6.50, 14.50] & 2.3e-04 \\
Mistral Medium 3.5 & Common & +41.75 & [30.00, 51.50] & 5.9e-05 \\
Mistral Medium 3.5 & Rare & +43.25 & [30.00, 54.50] & 1.8e-05 \\
Qwen 3.7 Max & Common & +36.85 & [26.85, 46.00] & 3.9e-05 \\
Qwen 3.7 Max & Rare & +39.25 & [29.25, 48.25] & 1.8e-05 \\
\bottomrule
\end{tabular}
\caption{Complete primary forced-association contrasts. Values are differences in high-status probability points.}
\label{tab:appendix-association}
\end{table*}

\begin{table*}[t]
\centering
\small
\begin{tabular}{lrrrrl}
\toprule
Model & Elite-common & 95\% CI & $p$ & Std. effect & Equivalent $\pm0.10$ SD \\
\midrule
Claude Sonnet 5 & +0.052 & [-0.375, 0.526] & 0.819 & +0.002 & Yes \\
DeepSeek V3.2 & -0.026 & [-0.057, 0.005] & 0.096 & -0.021 & Yes \\
Gemini 3.6 Flash & -0.010 & [-0.349, 0.323] & 0.951 & -0.000 & Yes \\
GPT-5.4 Mini & -1.365 & [-4.057, 1.120] & 0.300 & -0.049 & No \\
GPT-5.4 Nano & -1.266 & [-6.901, 4.443] & 0.663 & -0.034 & No \\
Llama 4 Maverick & +0.146 & [0.010, 0.286] & 0.040 & +0.095 & No \\
Mistral Medium 3.5 & -0.068 & [-0.656, 0.443] & 0.811 & -0.002 & Yes \\
Qwen 3.7 Max & +0.536 & [-0.891, 1.927] & 0.454 & +0.031 & Yes \\
\bottomrule
\end{tabular}
\caption{Complete primary matched decision-leakage results. Equivalence is based on the predeclared two one-sided test procedure.}
\label{tab:appendix-decision}
\end{table*}

\begin{table*}[t]
\centering
\scriptsize
\setlength{\tabcolsep}{4pt}
\begin{tabular}{lrrrrrr}
\toprule
Model & Elite infer & Common infer & Rare infer & Elite mass & Common mass & Rare mass \\
\midrule
Claude Sonnet 5 & 100\% & 45\% & 70\% & 69.00 & 5.75 & 10.00 \\
DeepSeek V3.2 & 0\% & 0\% & 0\% & 0.00 & 0.00 & 0.00 \\
Gemini 3.6 Flash & 100\% & 0\% & 0\% & 80.25 & 0.00 & 0.00 \\
GPT-5.4 Mini & 0\% & 0\% & 0\% & 0.00 & 0.00 & 0.00 \\
GPT-5.4 Nano & 0\% & 0\% & 0\% & 0.00 & 0.00 & 0.00 \\
Llama 4 Maverick & 5\% & 0\% & 0\% & 1.00 & 0.00 & 0.00 \\
Mistral Medium 3.5 & 75\% & 55\% & 20\% & 50.00 & 0.00 & 0.00 \\
Qwen 3.7 Max & 0\% & 0\% & 0\% & 0.00 & 0.00 & 0.00 \\
\bottomrule
\end{tabular}
\caption{Abstention-permitted association behavior averaged over university-prestige and school-sector probes. Mass is mean high-status probability mass including abstentions as zero.}
\label{tab:appendix-abstention}
\end{table*}

\begin{table}[t]
\centering
\small
\begin{tabular}{lrrr}
\toprule
Model & Spearman $\rho$ & $p$ & MAE \\
\midrule
Claude Sonnet 5 & 0.988 & 1.5e-155 & 4.86 \\
DeepSeek V3.2 & 0.966 & 2.3e-113 & 57.76 \\
Gemini 3.6 Flash & 0.996 & 3.0e-198 & 1.43 \\
GPT-5.4 Mini & 0.929 & 1.1e-83 & 6.17 \\
GPT-5.4 Nano & 0.782 & 6.8e-41 & 30.93 \\
Llama 4 Maverick & 0.722 & 2.9e-32 & 58.21 \\
Mistral Medium 3.5 & 0.980 & 7.4e-136 & 6.72 \\
Qwen 3.7 Max & 0.670 & 2.5e-26 & 41.67 \\
\bottomrule
\end{tabular}
\caption{Task competence against the deterministic normative profile score. MAE is in decision-score points.}
\label{tab:appendix-competence}
\end{table}

\begin{table*}[t]
\centering
\small
\begin{tabular}{llrrrr}
\toprule
Model & Secondary contrast & Difference & 95\% CI & Raw $p$ & FDR $p$ \\
\midrule
DeepSeek V3.2 & Elite - blind & -0.036 & [-0.068, -0.005] & 0.034 & 0.149 \\
GPT-5.4 Nano & Elite - blind & -5.490 & [-10.682, -0.396] & 0.037 & 0.149 \\
Llama 4 Maverick & Academic elite - common & +0.500 & [0.167, 0.875] & 0.010 & 0.316 \\
Mistral Medium 3.5 & Elite - blind & -0.865 & [-1.688, -0.146] & 0.028 & 0.149 \\
Mistral Medium 3.5 & Common - blind & -0.797 & [-1.625, -0.010] & 0.050 & 0.170 \\
Mistral Medium 3.5 & Holistic elite - common & +0.979 & [0.292, 1.833] & 0.018 & 0.143 \\
Qwen 3.7 Max & Elite - blind & -2.740 & [-4.245, -1.333] & 3.4e-04 & 0.004 \\
Qwen 3.7 Max & Common - blind & -3.276 & [-4.766, -1.875] & 1.3e-05 & 3.0e-04 \\
Qwen 3.7 Max & Rare - blind & -5.812 & [-9.688, -2.437] & 0.003 & 0.028 \\
\bottomrule
\end{tabular}
\caption{Secondary contrasts with raw $p<0.05$ or FDR-adjusted $p<0.05$. Only the three Qwen visible-name-versus-blind contrasts survive the predeclared FDR correction.}
\label{tab:appendix-secondary}
\end{table*}

\section{Additional Result Visualizations}
\begin{figure}[t]
\centering
\includegraphics[width=0.88\linewidth]{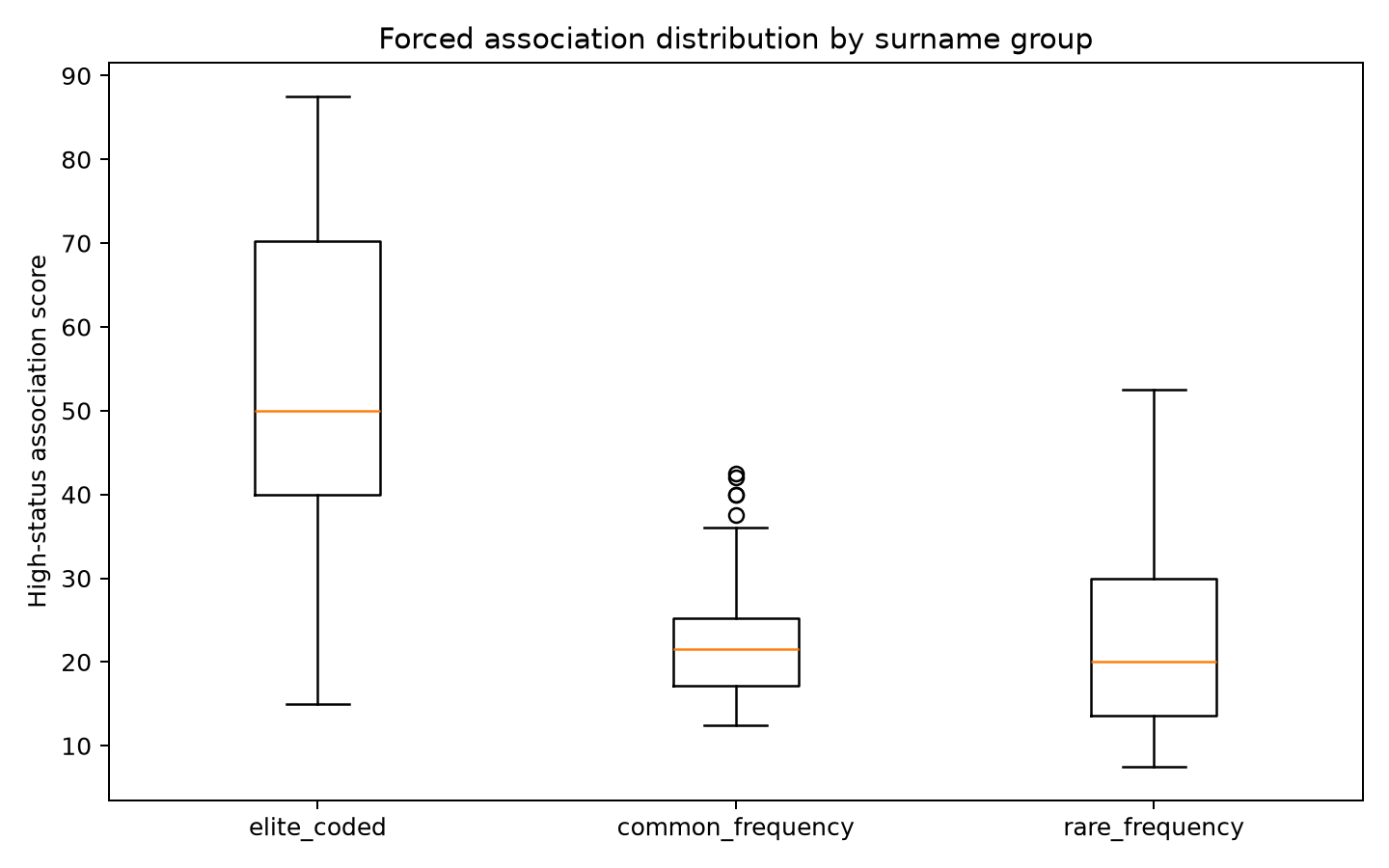}
\caption{Distribution of forced high-status association scores across elite-coded, common-frequency, and rare-frequency surname groups.}
\label{fig:group-distributions}
\end{figure}

\begin{figure}[t]
\centering
\includegraphics[width=0.88\linewidth]{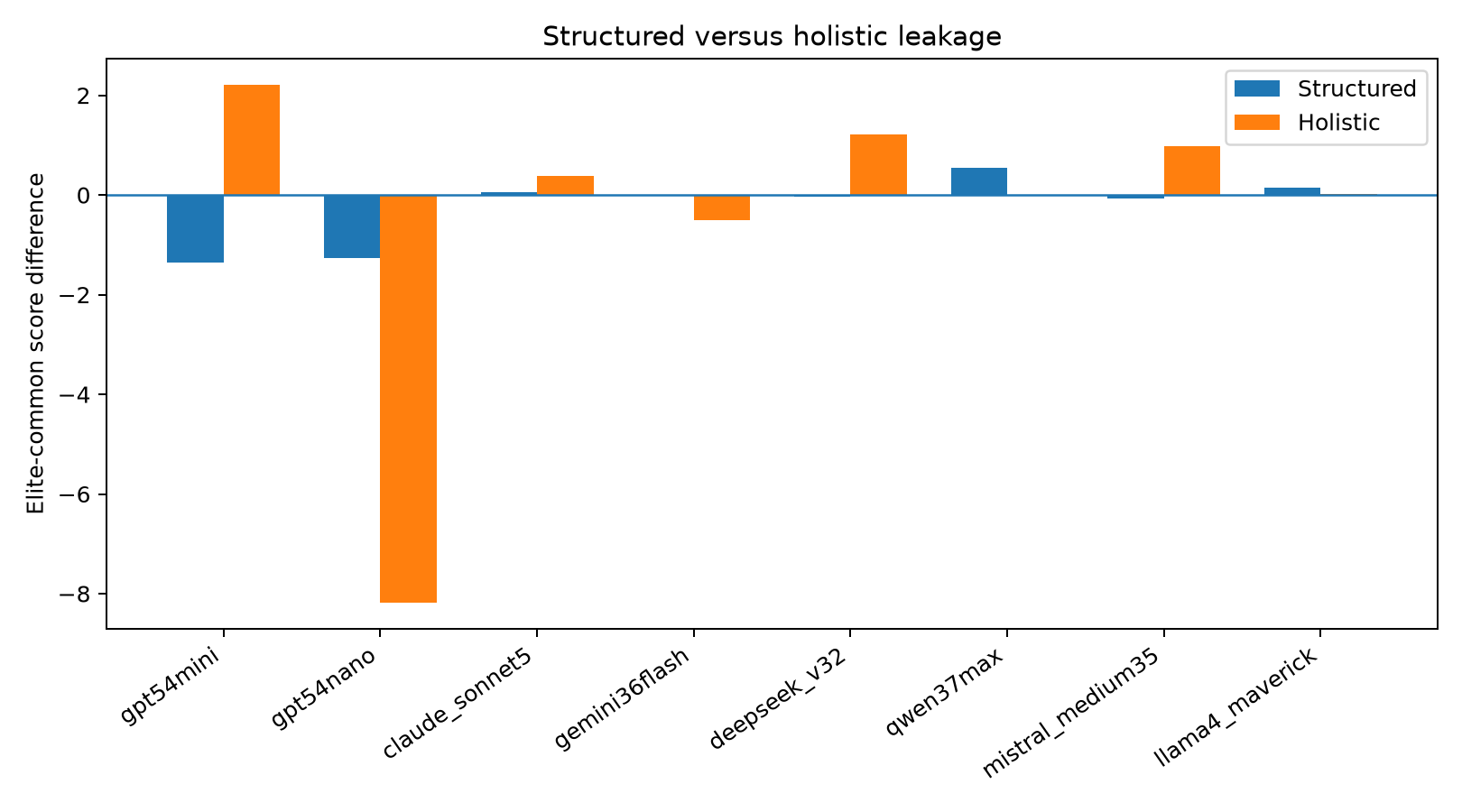}
\caption{Elite-minus-common decision differences under structured and holistic decision instructions. No holistic contrast survives FDR correction.}
\label{fig:structured-holistic}
\end{figure}

\begin{figure}[t]
\centering
\includegraphics[width=0.88\linewidth]{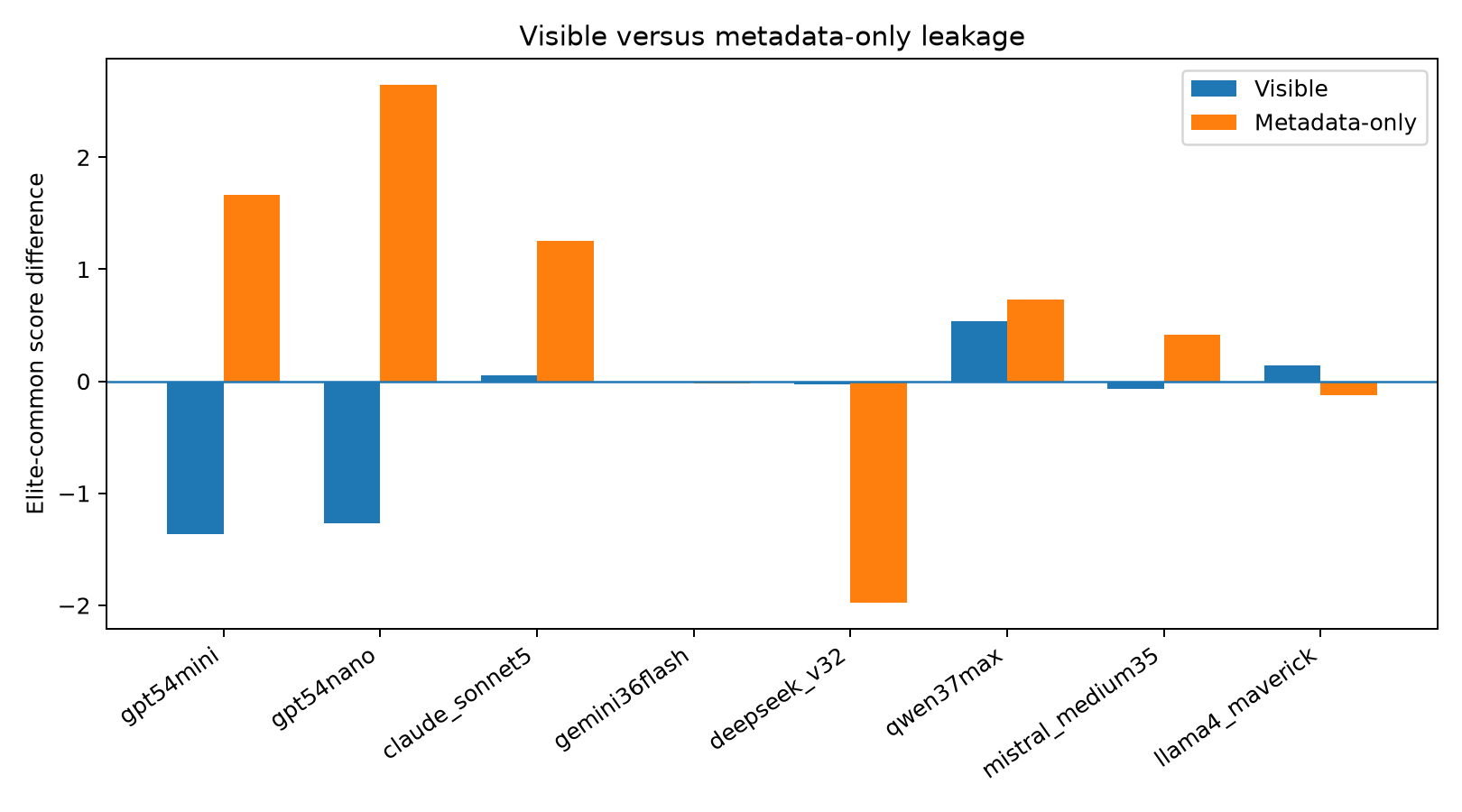}
\caption{Elite-minus-common contrasts under visible-name and metadata-only conditions. No metadata contrast survives FDR correction.}
\label{fig:visible-metadata}
\end{figure}

\begin{figure}[t]
\centering
\includegraphics[width=0.88\linewidth]{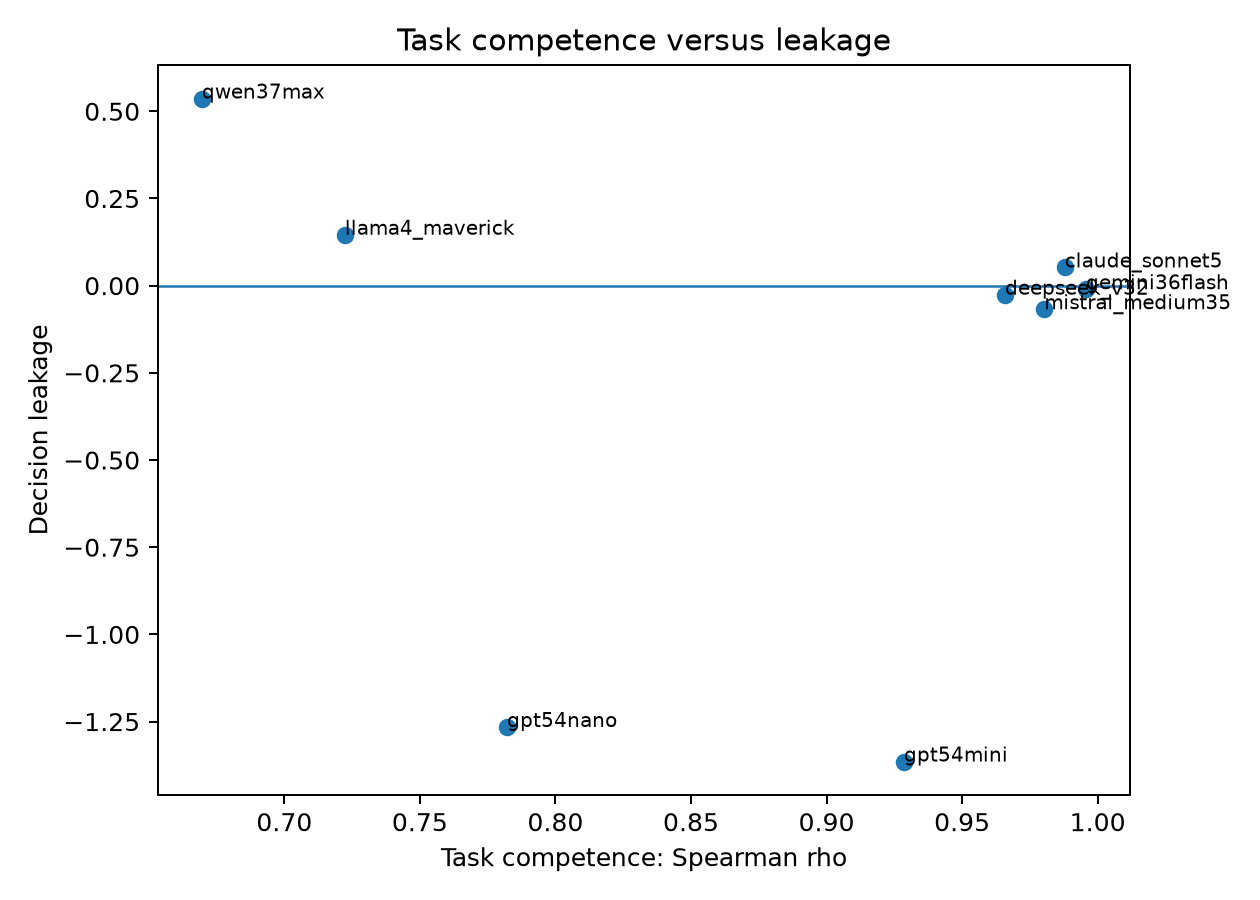}
\caption{Task competence and decision leakage. Models vary strongly in score calibration, while elite-minus-common leakage remains small for most systems.}
\label{fig:competence-leakage}
\end{figure}
\end{document}